\documentclass[aps,pre,preprint,superscriptaddress,showpacs,amssymb]{revtex4}
\usepackage{graphicx}
\usepackage{graphics}
\usepackage{amsmath}
\usepackage{tabularx}
\usepackage{color}
\usepackage{doi}
\usepackage{textcomp}

\begin{document}

\title{Asynchronous SIR model on Two-Dimensional Quasiperiodic Lattices}

\author{G. B. M. Santos}
\affiliation{Departamento de F\'{\i}sica, Universidade Federal do Piau\'{i}, 57072-970, Teresina - PI, Brazil}
\author{T. F. A. Alves}
\affiliation{Departamento de F\'{\i}sica, Universidade Federal do Piau\'{i}, 57072-970, Teresina - PI, Brazil}
\author{G. A. Alves}
\affiliation{Departamento de F\'{i}sica, Universidade Estadual do Piau\'{i}, 64002-150, Teresina - PI, Brazil}
\author{A. Macedo-Filho}
\affiliation{Campus Prof.\ Antonio Geovanne Alves de Sousa, Universidade Estadual do Piau\'i, 64260-000, Piripiri - PI, Brazil}

\date{Received: date / Revised version: date}

\begin{abstract}

We considered the Asynchronous SIR (susceptible-infected-removed) model on Penrose and Ammann-Beenker quasiperiodic lattices, and obtained its critical behavior by using Newman-Ziff algorithm to track cluster propagation by making a tree structure of clusters grown at the dynamics, allowing to simulate SIR model on non-periodic lattices and measure any observable related to percolation. We numerically calculated the order parameter, defined in a geographical fashion by distinguish between an epidemic state, characterized by a spanning cluster formed by the removed nodes and the endemic state, where there is no spanning cluster. We obtained the averaged mean cluster size which plays the role of a susceptibility, and a cumulant ratio defined for percolation to estimate the epidemic threshold. Our numerical results suggest that the system falls into two-dimensional dynamic percolation universality class and the quasiperiodic order is irrelevant, in according to results for classical percolation.

\end{abstract}


\maketitle

\section{Introduction}

Asynchronous SIR (susceptible-infected-removed) model\cite{Bernoulli-1760, Othsuki-1986, Renshaw-1991, Keeling-2007, Henkel2008, Tome-2010, deSouza-2011, Tome-2011, Tome-2015, Ruziska-2017, Pastorsatorras2015} is a stochastic model meant to simulate epidemic outbreaks. The essential ingredients to accomplish that are all present in the model: first, it has the needed spatial structure, composed of a lattice with bonds which can represent the contacts between individuals, and second, it has a dynamics composed of stoichiometric equations describing temporal evolution of a disease. Stochasticity itself is the third ingredient, which introduces fluctuations. Fluctuations are an important distinction between stochastic models and deterministic ones to simulate a population of individuals where a disease can spread\cite{Renshaw-1991, Dickman1999, Keeling-2007, Tome-2015}.

SIR model is one of the simplest models of epidemic spreading by contact, where infected subjects can recover and gain permanent immunity (or die). Another simple model of disease spreading is the SIS (susceptible-infected-susceptible) model (and its variant, the Contact Process model)\cite{Pastorsatorras2015, Pastorsatorras2013, TEHarris1974, Castellano2006, Pastorsatorras2016, Jensen1993, Lubeck2002, Lubeck2003, Jansen2007, Henkel2008, Oliveira2008, Silva2013, Almeida2016, Alves-2018a, Alves-2018b, Schrauth-2018}. The main difference between SIR and SIS models is the permanent immunity (or removal) present in SIR model. This important distinction is the cause of these two models fall into distinct universality classes: dynamical percolation\cite{Othsuki-1986, Pastorsatorras2015, Tome-2010, deSouza-2011} and directed percolation\cite{Janssen1981, Grassberger1982} for SIR and SIS models, respectively. SIR model is appropriate to simulate the spreading diseases like Ebola, SARS, AIDS, Chickenpox, etc., where death/removal or permanent immunity are one of the states an individual can assume.

It is known that SIR model can display a continuous phase transition between an endemic state to an epidemic state\cite{Tome-2010, deSouza-2011}, where epidemic and endemic phases are defined in a geographical fashion. The epidemic phase is characterized by the presence of an ``infinite component'', identified by a percolating cluster formed by the activated sites (eventually removed ones at the end of dynamics). The critical threshold between the epidemic (percolating) and endemic (non-percolating) phases is controlled by the infection and recover rates. The infection rate $\mu_{i}$ is the probability with an infected particle can successfully infect one of its neighbors in a Poisson process and the recover rate $\mu_{c}$ is the probability of an infected particle can gain permanent immunity (or ``die''), being excluded from the dynamics. Every path, starting from a ``patient zero'' (one randomly chosen individual to be the first infected one of the epidemic outbreak), subjecting to both infection and recover rates with permanent immunity, always evolve to an absorbing state where there are not any infected particles. Presence of absorbing states negates the detailed balance condition, therefore SIR model is a non equilibrium one.

In this present paper, we considered SIR model on Penrose and Ammann-Beenker quasiperiodic lattices. Our objectives are two-fold:
\begin{itemize}
\item First, we propose for the first time a simple way to simulate SIR model on non-periodic lattices by using Newman-Ziff algorithm\cite{Newman-2001} to track clusters generated by the dynamics and to obtain percolation observables like the mean cluster size;
\item Second, we investigate if quasiperiodic order can change the SIR model critical behavior. There are some criteria trying to predict if quenched disorder, or quasiperiodic order, can change the critical exponents of a given model, like Harris\cite{ABHarris1974}, or Harris-Luck\cite{Luck1993} criteria, respectively. Both Harris and Harris-Luck criteria are superseded by Harris-Barghathi-Vojta criterion\cite{Vojta2009,Barghathi2014}, however, even Harris-Barghathi-Vojta criterion is known to fail\cite{Schrauth-2018}.
\end{itemize}
Therefore, to find if quasiperiodic order can change the universality class of SIR model, we obtained the relevant parameters: the order parameter, defined as the average percolating cluster density, the average mean cluster size, and Binder cumulant defined for percolation in order to estimate the epidemic threshold and critical exponents by using standard finite size scaling (FSS) techniques. To correctly determine the universality class, we estimated three independent critical exponent ratios: $1/\nu$, $\beta/\nu$ and $\gamma/\nu$.

This paper is organized as follows: in section II we describe the SIR model and its dynamics, lattice building algorithm and details of calculated parameters by Newman-Ziff algorithm, in section III we discuss our numerical results and in section IV we present our conclusions.

\section{Model and Implementation}

We considered the Asynchronous SIR model, coupled to the two-dimensional (2D) Penrose and Ammann-Beenker quasiperiodic lattices. SIR model is a compartmental model, where a population, composed of vertexes in a lattice, is divided in three compartments, corresponding to three possible states, namely:
\begin{itemize}
\item \textbf{Susceptible state}: Individuals in that state are not infected, are not infecting and are not immune/dead;
\item \textbf{Infected state}: Individuals in that state can infect another individuals, and are not immune/dead;
\item \textbf{Removed state}: Individuals in that state do not infect and are not infecting because of permanent immunity/being dead. They do not interact with individuals in the two other compartments.
\end{itemize}
We can associate the lattice state as a vector $\psi_i$ which can take the values $0$, $1$ and $-1$, corresponding to susceptible, infected and removed individuals, respectively.

SIR model can be expressed as a reaction-diffusion process, where individuals can change its state and go from one compartment to another\cite{vanKampen-1981}. Transitions of the individuals between compartments are represented by the following stoichiometric reactions
\begin{eqnarray}
&& S + I \xrightarrow{\mu_i} 2I \nonumber \\
&& I \xrightarrow{\mu_c} R.
\label{sir-estoichometric}
\end{eqnarray}
The stoichiometric equations (\ref{sir-estoichometric}) can be mapped in a lattice Markovian dynamics with the following rules:
\begin{itemize}
\item [(1)] We start with only one infected vertex, the ``patient zero'', randomly chosen from the lattice with a population of $N$ individuals, attached to the respective lattice vertexes. Along the entire dynamics, two lists are updated: a list of infected individuals and a list of removed ones. The infected list starts with the ``patient zero'' and the list of removed sites begins empty.
\item [(2)] Next, we update the system state by randomly choosing an infected vertex $i$ from the infected list and proceed as follows:
	\begin{itemize}
	\item[(a)] A random uniform number $x$ in the interval $[0,1)$ is generated. If $x<\lambda$, the infected site is removed from the infected list and placed in the removed list;
	\item[(b)] If $x>\lambda$, we pick a neighboring vertex randomly to add it on the infected list;
	\end{itemize}
\item[(3)] The step (2) is repeated several times until the system have no infected sites, i.e., reach any absorbing state. For each pass of the dynamics we can increment the dynamics clock by a time interval $1/N_i$ where $N_i$ is the number of infected particles. Note that the update is made in an asynchronous way.
\end{itemize}
From the above kinetic Monte Carlo rules, we can see that the recover and infection rates are given by $\mu_{c}=\lambda$, and $\mu_{i}=1-\lambda$, respectively.

Some observables can be obtained at the absorbing state, for example, the number of removed vertexes $N_r$ as function of $\lambda$, expressed as
\begin{equation}
N_r = \sum_{i=1}^{n}\left| \psi_i \right|.
\end{equation}
However, just $N_r$ is not sufficient to fully determine the critical behavior because it is not the order parameter. Following classical percolation theory\cite{Stauffer-1992,Christensen-2005}, we need the observables related to the clusters formed by the removed sites and to accomplish that, we have to determine the cluster distribution $n_{\mathrm{cluster}}(s)$, i.e., the number of clusters with $s$ removed vertexes. The cluster distribution can be measured by using Newman-Ziff algorithm\cite{Newman-2001} with the feature of identifying if there is a percolating cluster as a result of the dynamics. What is a percolating cluster depends on the boundary conditions: in the case of periodic boundaries, it is a wrapping cluster; for non periodic boundaries, a percolating cluster is a spanning cluster\cite{Newman-2001}.

From the cluster distribution, we have the fraction of removed vertexes in the finite (non percolating) clusters with size $s$
\begin{equation}
P_s = s n_{\mathrm{cluster}}(s).
\label{fraction-cluster}
\end{equation}
The following identity is then obvious
\begin{equation}
p = P_\infty + \sum_s s n_{\mathrm{cluster}}(s),
\end{equation}
where $p$ is the total fraction of removed vertexes and $P_\infty$ is the fraction of removed vertexes in the percolating cluster. The epidemic phase is then defined in a geographical way, where we should have a percolating cluster emulating an epidemic spreading reaching the most remote points of a lattice. At the epidemic threshold, size distribution of clusters should be a power law\cite{Tome-2010}, i.e.
\begin{equation}
n_{\mathrm{cluster}}(s,\lambda_{c}) \propto s^{-\tau},
\label{cluster-distrib}
\end{equation}
and, in general, we have the following scaling \textit{ansatz} in the vicinity of the epidemic threshold
\begin{equation}
n_{\mathrm{cluster}}(s) = s^{-\tau}\mathrm{F}\left[ \left( \lambda-\lambda_c\right)s^\sigma \right],
\label{scaling-ansatz}
\end{equation}
where $\mathrm{F}$ is a fast decaying function\cite{Stauffer-1992,Christensen-2005}.

The epidemic outbreak should result in a percolating cluster at the epidemic phase, i.e. $P_\infty \ne 0$. Therefore, the order parameter is the percolating cluster density
\begin{equation}
P = \left< P_\infty \right> 
\label{orderparameter}
\end{equation}
where the average is done over dynamics realizations. Another relevant parameter is the mean cluster size $S$
\begin{equation}
S = \frac{1}{N_r}\sum_s s P_s =  \frac{1}{N_r}\sum_s s^2 n_{\mathrm{cluster}}(s),
\label{mean-cluster-size-star}
\end{equation}
The mean cluster size $S$ plays the role of the susceptibility in classical percolation theory, which is defined by the following average
\begin{equation}
\chi = \left< S \right>.
\label{susceptibility}
\end{equation}

We can define another two observables: first is the mean cluster size $S$ where the summation now includes the percolating cluster, with size $s_\mathrm{perc}$
\begin{equation}
S' = \frac{1}{N_r}\left(s^2_\mathrm{perc} + \sum_s s^2 n_{\mathrm{cluster}}(s)\right),
\label{mean-cluster-size}
\end{equation}
and the second is the mean quadratic cluster size
\begin{equation}
M' = \frac{1}{N_r}\left(s^3_\mathrm{perc} + \sum_s s^3 n_{\mathrm{cluster}}(s)\right).
\label{mean-quadratic-cluster-size}
\end{equation}
In the infinite size lattice limit, $s_\mathrm{perc} \rightarrow \infty$ and in the endemic phase, $s_{perc}=0$, therefore observables $S'$ and $M'$ only make sense for finite lattice sizes. They are needed to express a quantity analogous to Binder cumulants for percolation\cite{deSouza-2011} as follows.

From the scaling \textit{ansatz} written in Eq.(\ref{scaling-ansatz}), the relevant observables in Eqs. (\ref{orderparameter}), (\ref{susceptibility}), (\ref{mean-cluster-size}), and (\ref{mean-quadratic-cluster-size}) should scale as
\begin{eqnarray}
P &\approx& L^{-\beta/\nu}f_{P}\left( L^{1/\nu} \left| \lambda - \lambda_c \right| \right), \label{orderparameter-fss} \\
\chi &\approx& L^{\gamma/\nu}f_\chi\left( L^{1/\nu} \left| \lambda - \lambda_c \right| \right), \label{susceptibility-fss} \\
\left< S' \right> &\approx & L^{\gamma/\nu}f_{\left< S' \right>}\left( L^{1/\nu} \left| \lambda - \lambda_c \right| \right), \label{sobservable-fss} \\
\left< M' \right> &\approx & L^{\left(\beta+2\gamma\right)/\nu}f_{\left< M' \right>}\left( L^{1/\nu} \left| \lambda - \lambda_c \right| \right), \label{mobservable-fss} 
\end{eqnarray}
according to classical percolation theory\cite{Stauffer-1992,Christensen-2005}. Combining Eqs. (\ref{orderparameter-fss}), (\ref{sobservable-fss}), and (\ref{mobservable-fss}), we find that the quantity
\begin{equation}
U = P \frac{\left< M' \right>}{\left< S' \right>^2}
\label{bindercumulant}
\end{equation}
should be universal at the epidemic threshold where the scaling dependencies cancel out, being analogous to the Binder cumulant for ferromagnetic spin models\cite{deSouza-2011}. Therefore, $U$ obeys the following scaling
\begin{equation}
U \approx f_U\left( L^{1/\nu} \left| \lambda - \lambda_c \right| \right).
\label{bindercumulant-fss}
\end{equation}
Binder cumulant curve crossings for different lattice sizes give an estimate for the epidemic threshold $\lambda_c$, and scaling relations on Eqs. (\ref{orderparameter-fss}), (\ref{susceptibility-fss}), and (\ref{bindercumulant-fss}) can be used to estimate the critical exponents $\beta$, $\gamma$ and $\nu$ by standard data collapses, respectively, to completely determine the universality class.

\begin{figure}[h]
\begin{center}
\includegraphics[scale=0.25]{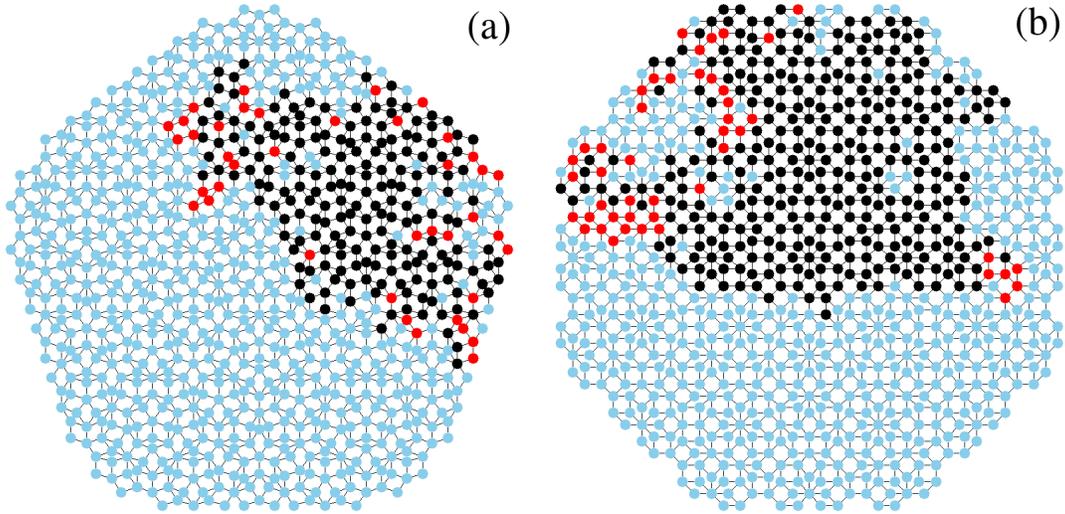} 
\end{center}
\caption{(Color Online) Clusters resulting from SIR model dynamics in (a) Penrose lattice and (b) Ammann-Beenker lattice. In light grey (blue), darker grey (red) and black, we have susceptible, infected and removed vertexes, respectively. Note that the activity is restricted to cluster boundaries in a way that only one cluster is grown as a result of the dynamics.}
\label{cluster-fig}
\end{figure}

Concerning the particular case of SIR model, only one cluster of size $s=N_r$ is grown for each realization of the dynamics. Typical clusters are shown in Fig.(\ref{cluster-fig}). In this way, we have for SIR model at the endemic phase $n_{\mathrm{cluster}}(s)=\delta_{s,N_r}$ and $s_\mathrm{perc}=0$, where $\delta$ is a Kronecker index, and for the epidemic (percolating) phase we should have $n_{\mathrm{cluster}}(s)=0$ and $s_\mathrm{perc}=N_r$. Therefore $S'=N_r$ and $M'=N_r^2$. On the other hand, the order parameter reduces to $P_\infty=0$ if there is not a percolating cluster and to $P_\infty=1$ for the converse. One should note that this particular simpler cluster structure can be simulated without Newman-Ziff algorithm\cite{deSouza-2011}, but at doing this, we can lose the power to generalize the model for non-periodic and random lattices. Indeed, Newman-Ziff algorithm can be applied with modifications to identify a spanning cluster in the case of non-periodic boundaries or a wrapping cluster, in the case of periodic boundaries\cite{Yang-2012}.

\begin{figure}[h!]
\begin{center}
\includegraphics[scale=0.2]{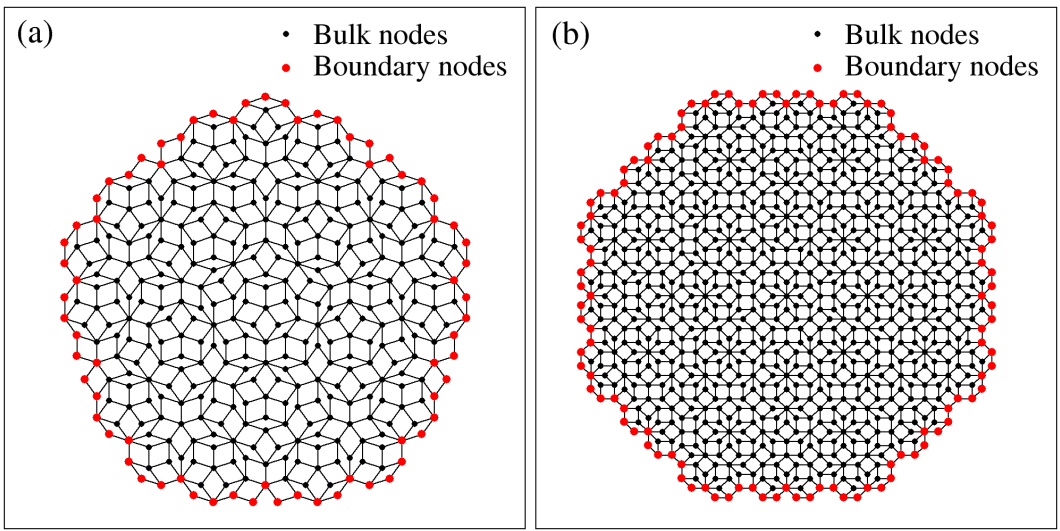}
\end{center}
\caption{(Color Online) Third generation ($g=3$) of (a) Penrose and (b) Ammann-Beenker lattices. In black we have the bulk nodes and in grey (red) we have the boundary nodes.}
\label{latticegenerations}
\end{figure}

Regarding the lattice structure, Penrose and Ammann-Beenker lattices were generated by projection algorithm\cite{deBruijn1981, Naumis2003, Alves-2018b}, where we can define a generation parameter $g$ for these lattices. The number of lattice vertexes is a function of $g$. We depict Penrose and Ammann-Beenker quasiperiodic lattices for $g=3$ in Figs.(\ref{latticegenerations}a) and (\ref{latticegenerations}b), respectively. Note that the boundary nodes are all identified, which are needed to determine if there is a spanning cluster.

To correctly identify all the boundary nodes, one should search by four classes of vertexes in the following order
\begin{itemize}
\item[1:] Vertexes with only two neighbors;
\item[2:] The two neighbors of class 1 vertexes;
\item[3:] In the case of two neighboring boundary rhombi sharing a vertex, the shared vertex is a boundary one;
\item[4:] In the case of two neighboring boundary rhombi sharing an edge, one of two nodes shared is a boundary one. This node have exactly three neighbors, all contained on boundary rhombi. If the two shared edge vertexes have three neighbors, the vertex located farther from the central lattice node is the boundary node.
\end{itemize}
Boundary rhombi can be identified by looking for extremal grid points in grid lines when calculate the multigrid crossings\cite{Alves-2018b}. Spanning clusters are defined in a different way for Penrose and Ammann-Beenker lattices because of rotational symmetry:
\begin{itemize}
\item For Penrose lattice, a spanning cluster is a cluster who contains any pair of vertexes symmetrical over a lattice rotation of $\pm 4\pi/5$ about the central node;
\item For Ammann-Beenker lattice, a spanning cluster is a cluster who contains any pair of vertexes symmetrical over a lattice rotation of $\pi$ about the central node. 
\end{itemize}
Note that we used non periodic boundaries and our results should be prone to finite size boundary effects, but they can be avoided by just using bigger lattice sizes as expected for amenable graphs. Generations from $g=4$ to $g=10$, used in our numerical results, give good data collapses for both lattices, indicating that these sizes are enough to avoid boundary effects.

We repeated SIR model dynamics $10^{6}$ times for every control parameter $\lambda$, starting from ``patient zero'' until the system evolved to an absorbing state, growing $10^{6}$ clusters. We calculated an ensemble composed of $10^{6}$ measurements of order parameter, mean cluster size and Binder cumulant, one for each cluster generated, to take the relevant ensemble averages. We calculated statistical errors by using the ``jackknife'' resampling\cite{Tukey1958}. In the following section, we discuss our obtained numerical results.

\section{Results and Discussion}

\begin{figure}[p]
\begin{center}
\includegraphics[scale=0.17]{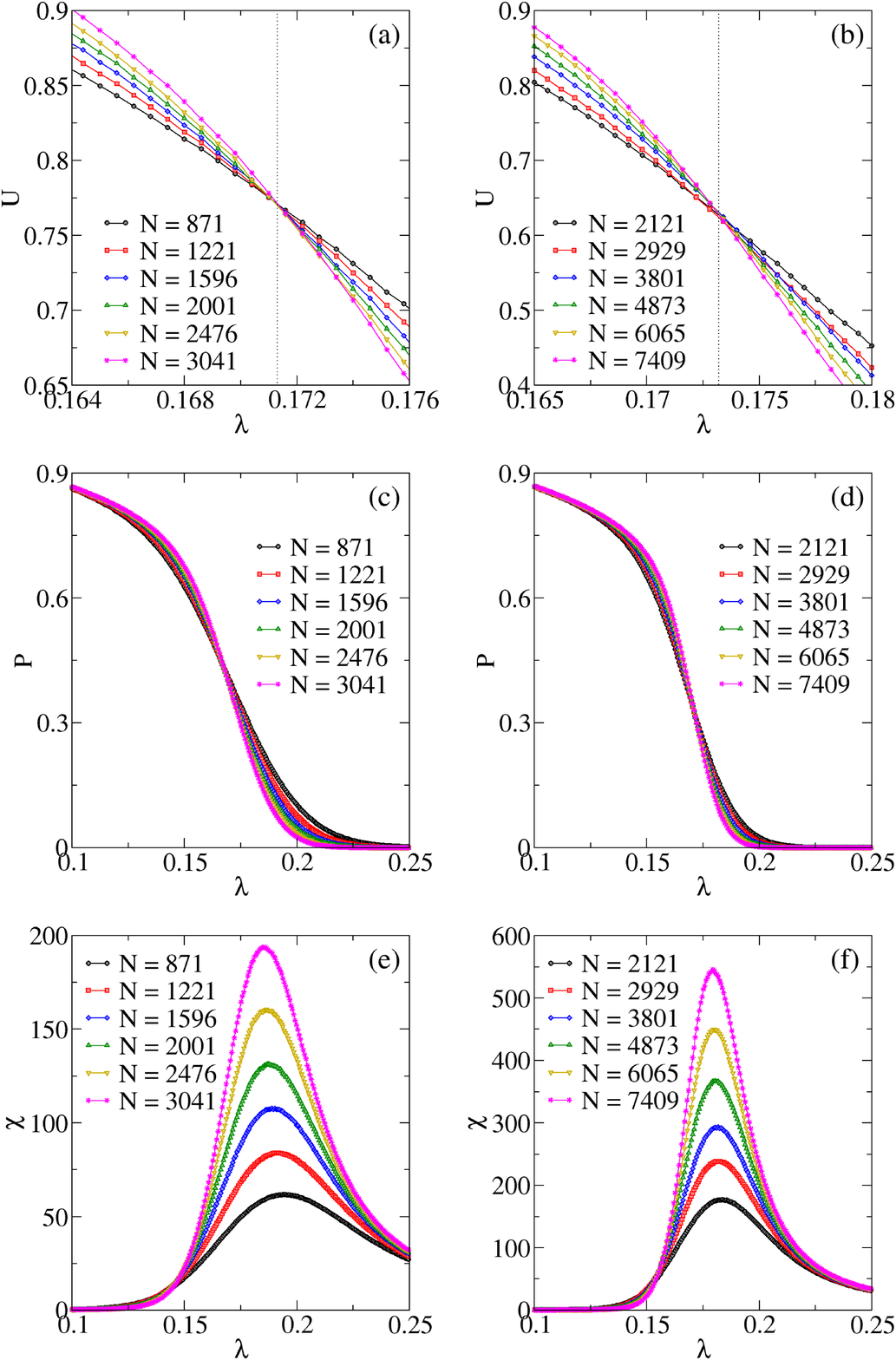}
\end{center}
\caption{In panels (a), (c), and (e), we show our numerical data for the Binder cumulant $U$, average spanning cluster density $P$ and average mean cluster size $\chi$ as functions of recovery rate $\lambda$ for Penrose lattice, respectively. The same in panels (b), (d), and (f) for Ammann-Beenker lattice. Both lattices display a continuous phase transition. The epidemic thresholds are $\lambda_c = 0.1713(2)$ and $\lambda_c = 0.1732(5)$ for Penrose and Ammann-Beenker lattices, respectively. Statistical errors are smaller than symbols.}
\label{observables-fig}
\end{figure}

In this section, we show our numerical results of Asynchronous SIR model coupled to Penrose and Ammann-Beenker quasiperiodic lattices. First, we show results for the Binder cumulant in Figs.(\ref{observables-fig}a) and (\ref{observables-fig}b). From Binder cumulant crossings, we estimated the epidemic thresholds at $\lambda_c = 0.1713(2)$ for Penrose lattice and $\lambda_c = 0.1732(5)$ for Ammann-Beenker lattice. Binder cumulant expressed in Eq.(\ref{bindercumulant}) should not depend on the lattice size at the critical point, but it should depend on the dimensionality and boundary conditions as seen from comparing Figs.(\ref{observables-fig}a) and (\ref{observables-fig}b). Note that the Binder cumulant values at the critical threshold are not the same for Penrose and Ammann-Beenker lattices because of different boundary conditions we adopted to identify spanning clusters.

Following Binder cumulant discussion, we show order parameter results in Figs.(\ref{observables-fig}c) and (\ref{observables-fig}d), i.e, results for the average spanning cluster density. We see that the curves for both Penrose and Ammann-Beenker lattices have the typical sigmoidal shape, indicating a continuous phase transition from the epidemic phase to the endemic phase, where the order parameter vanishes, i.e., where we do not have any spanning cluster. By using Newman-Ziff algorithm, we were able to numerically calculate the averaged mean cluster size. In Figs.(\ref{observables-fig}e) and (\ref{observables-fig}f), we show the susceptibilities for Penrose and Ammann-Beenker lattices, respectively. They diverge at the epidemic threshold $\lambda_c$ in the infinite size lattice limit, according to the finite size scaling relation
\begin{equation}
\lambda_\mathrm{max} = \lambda_c + bL^{1/\nu},
\end{equation}
which means that the averaged mean cluster size maxima $\lambda_\mathrm{max}$ gets closer to the epidemic threshold $\lambda_c$ when increasing $N$, as expected for a continuous phase transition and as observed from our data.

\begin{figure}[p]
\begin{center}
\includegraphics[scale=0.17]{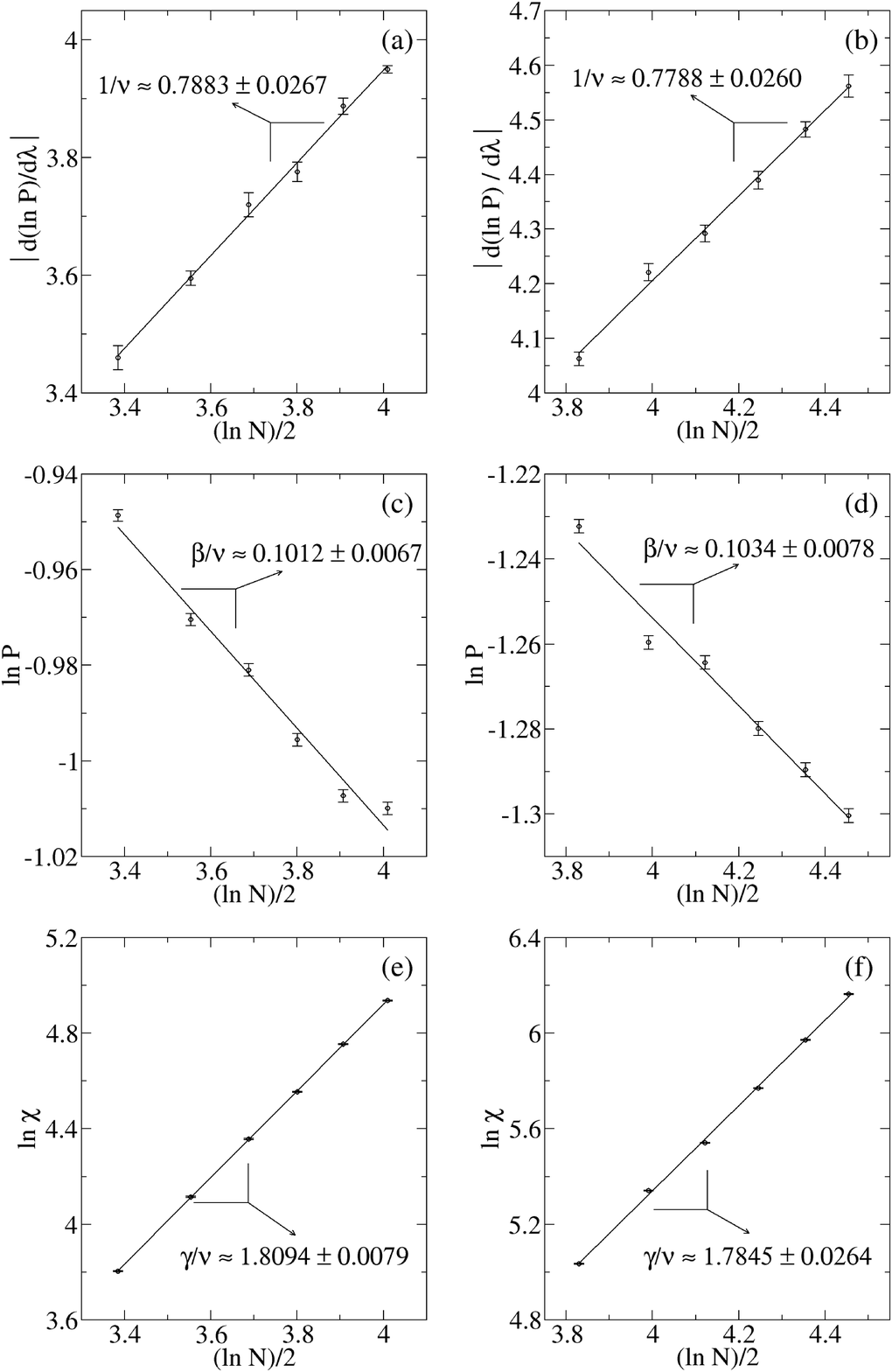}
\end{center}
\caption{In panels (a), (c), and (e), we show estimates of critical exponent ratios $1/\nu$, $\beta/\nu$, and $\gamma/\nu$ from finite size scaling regressions of $\left|\frac{\mathrm{d}\left(\ln{P}\right)}{\mathrm{d}\lambda}\right|$, $\ln P$, and $\ln \chi$, respectively, at the epidemic threshold $\lambda_c = 0.1713(2)$ as functions of $\left(\ln N \right)/2$ for Penrose lattice. The same for panels (b), (d), and (f) at the epidemic threshold $\lambda_c = 0.1732(5)$ for Ammann-Beenker lattice. All estimated exponent ratios are close to the respective 2D dynamic percolation exponents $1/\nu=3/4$, $\beta/\nu=5/48$, and $\gamma/\nu=43/24$. Note the logarithm scale for the data and statistical errors.}
\label{regressions-fig}
\end{figure}

Now, we discuss the critical exponent values. We estimated the critical exponent ratios $1/\nu$, $\beta/\nu$, and $\gamma/\nu$ by using finite size scaling regressions of $\left| \frac{\mathrm{d}\left(\ln{P}\right)}{\mathrm{d}\lambda}\right|$, $\ln P$, and $\ln \chi$, evaluated at the critical threshold $\lambda_c$, respectively. To extract $\left|\frac{\mathrm{d}\left(\ln{P}\right)}{\mathrm{d}\lambda}\right|$ data, we made linear regressions of the logarithm of order parameter in the vicinity of the epidemic threshold. Error bars are extracted in this case from the least squares method. In Figs.(\ref{regressions-fig}a) and Figs.(\ref{regressions-fig}b) we show data for $\left|\frac{\mathrm{d}\left(\ln{P}\right)}{\mathrm{d}\lambda}\right|$ as function of $\left(\ln N \right)/2$, allowing us to estimate $1/\nu \approx 0.7883 \pm 0.0267$ and $1/\nu \approx 0.7788 \pm 0.0260$ for Penrose and Ammann-Beenker lattices, respectively. These numerical values deviate $5\%$ and $4\%$, respectively, from the exact critical exponent ratio $1/\nu=3/4$ of 2D dynamical percolation.

Continuing the discussion of critical exponent ratios, in Figs.(\ref{regressions-fig}c) and (\ref{regressions-fig}d) we show the order parameter logarithm at the epidemic threshold as function of $\left(\ln N \right)/2$. In this case, finite size regressions should give estimates of $\beta/\nu$ critical exponent ratio, in a way we obtained $\beta/\nu \approx 0.1012 \pm 0.0067$ and $\beta/\nu \approx 0.1034 \pm 0.0078$ for Penrose and Ammann-Beenker lattices, respectively. Our numerical values for $\beta/\nu$ deviate $3\%$ and $1\%$ from exact $\beta/\nu$ value of 2D dynamical percolation universality class. We show the same for averaged mean cluster size in Figs.(\ref{regressions-fig}e) and (\ref{regressions-fig}f) at the epidemic threshold as function of $\left(\ln N \right)/2$. In an analogous way of the order parameter, we obtained $\gamma/\nu \approx 1.8094 \pm 0.0079$ and $\gamma/\nu \approx 1.7845 \pm 0.0264$ deviating $1\%$ and $0.5\%$ from the exact critical exponent ratio $\gamma/\nu=43/24$ for Penrose and Ammann-Beenker lattices, respectively.

\begin{figure}[p]
\begin{center}
\includegraphics[scale=0.17]{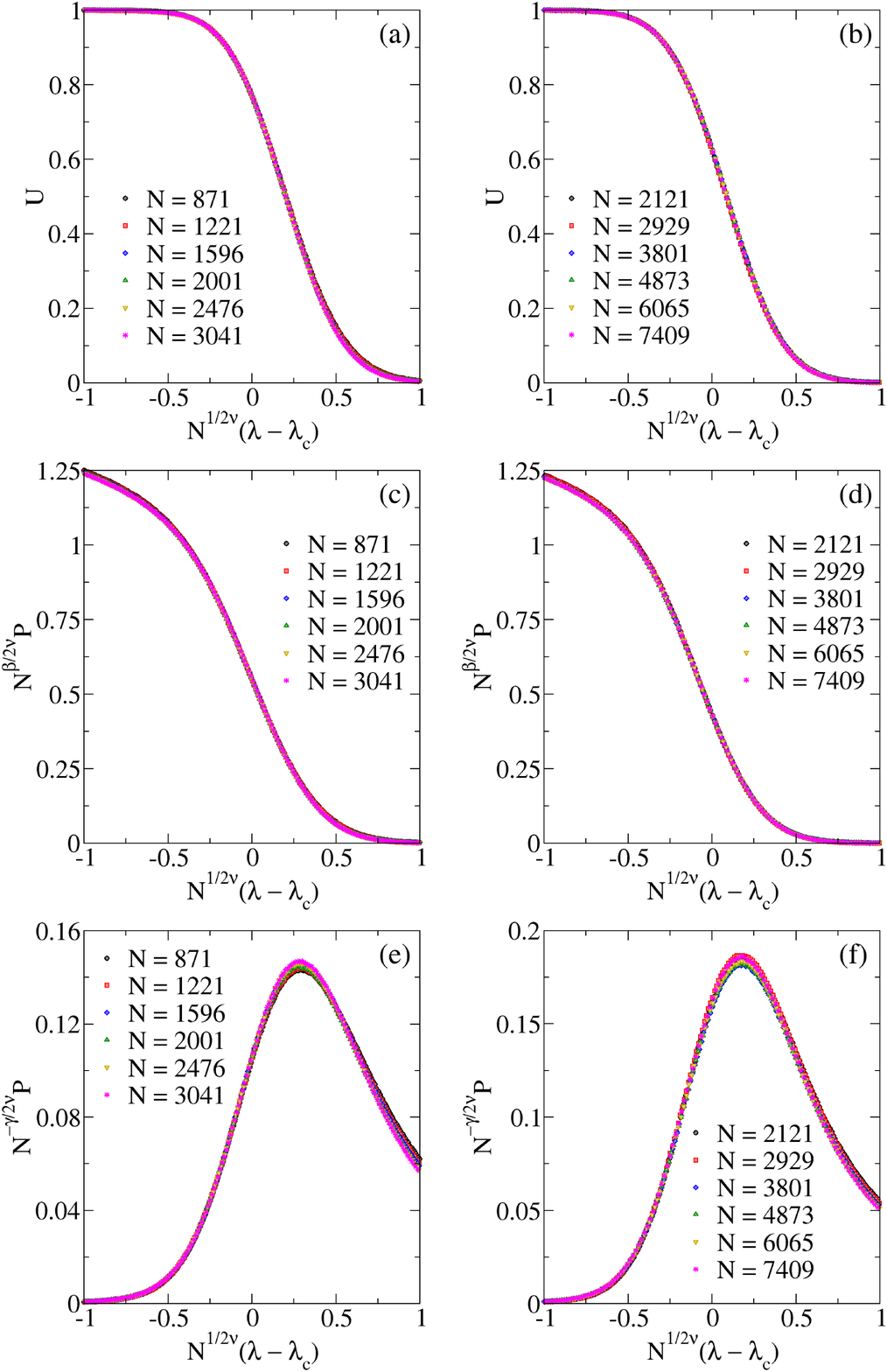}
\end{center}
\caption{In panels (a), (c), and (e), we show our numerical data for the Binder cumulant $U$, order parameter $P$ and susceptibility $\chi$ rescaled by using the finite size scaling relations on Eqs. (\ref{orderparameter-fss}), (\ref{susceptibility-fss}), and (\ref{bindercumulant-fss}) with the 2D dynamic percolation exponent ratios $1/\nu=3/4$, $\beta/\nu=5/48$, and $\gamma/\nu=43/24$ at the vicinity of epidemic threshold $\lambda_c = 0.1713(2)$ for Penrose lattice, respectively. The same is shown in panels (b), (d), and (f) at the vicinity of the epidemic threshold $\lambda_c = 0.1732(5)$ for Ammann-Beenker lattice.  Statistical errors are smaller than symbols.}
\label{collapsed-fig}
\end{figure}

By noting that the numerical values obtained from the finite size scaling regressions are all close from exact 2D dynamical percolation values, we can expect that the system falls into 2D percolation universality class. To confirm that, we collapsed all our numerical data shown in Fig.(\ref{observables-fig}) by using the finite size scaling relations written in Eqs. (\ref{orderparameter-fss}), (\ref{susceptibility-fss}), and (\ref{bindercumulant-fss}) with the known exact values of 2D dynamic percolation exponent ratios $1/\nu=3/4$, $\beta/\nu=5/48$, and $\gamma/\nu=43/24$. All data collapse panels shown in Fig.(\ref{collapsed-fig}) are respective to the panels shown in Fig.(\ref{observables-fig}). The data collapses for both Penrose and Ammann-Beenker lattices are a strong evidence that the system falls into dynamic percolation universality class in the same way of 2D square lattice\cite{Tome-2010, deSouza-2011}. Therefore the quasiperiodic order is irrelevant, in agreement with results for classical percolation in the same quasiperiodic lattices\cite{Sakamoto-1989, Babalievski-1995, Ziff-1999}.

\section{Conclusions}

We proposed a new way to simulate the Asynchronous SIR model on non periodic lattices by using Newman-Ziff algorithm to numerically determine the cluster size distribution and calculate any observable related to percolation for the first time. We applied the Newman-Ziff algorithm to identify the epidemic phase, characterized by a spanning cluster and used this technique to investigate SIR model coupled to Penrose and Ammann-Beenker lattices, and our numerical results suggest that the quasiperiodic order is irrelevant and the system falls into the 2D dynamical percolation universality class in the same way of the Asyncronous SIR model on periodic lattices, in agreement with classical percolation results. In order to characterize the critical behavior, we obtained the order parameter, defined as the spanning cluster density, the averaged mean cluster size and the Binder cumulant ratio defined for percolation. We estimated the epidemic thresholds: $\lambda_c = 0.1713(2)$ and $\lambda_c = 0.1732(5)$ for Penrose and Ammann-Beenker lattices, respectively.

\section{Acknowledgments}

We would like to thank CNPq (Conselho Nacional de Desenvolvimento Cient\'{\i}fico e tecnol\'{o}gico), FUNCAP (Funda\c{c}\~{a}o Cearense de Apoio ao Desenvolvimento Cient\'{\i}fico e Tecnol\'{o}gico) and FAPEPI (Funda\c{c}\~{a}o de Amparo a Pesquisa do Estado do Piau\'{\i}) for the financial support. We acknowledge the use of Dietrich Stauffer Computational Physics Lab, Teresina, Brazil, and Laborat\'{o}rio de F\'{\i}sica Te\'{o}rica e Modelagem Computacional - LFTMC, Piripiri, Brazil, where the numerical simulations were performed.

\bibliography{textv1}

\end{document}